\documentclass[aps,prb,twocolumn,amssymb,amsmath,superscriptaddress]{revtex4-1}
\usepackage{times}
\usepackage{graphicx}
\usepackage{color}
\usepackage{dcolumn}
\usepackage{bm}
\usepackage[colorlinks=true, letterpaper=true, pdfstartview=FitV, linkcolor=blue, citecolor=blue, urlcolor=blue]{hyperref}
\usepackage{appendix}

\begin{document}

\title{Unified framework of the microscopic Landau-Lifshitz-Gilbert equation \\
and its application to Skyrmion dynamics}

\author{Fuming Xu$^\S$}
\affiliation{College of Physics and Optoelectronic Engineering, Shenzhen University, Shenzhen 518060, China}
\author{Gaoyang Li$^\S$}
\affiliation{College of Physics and Optoelectronic Engineering, Shenzhen University, Shenzhen 518060, China}
\author{Jian Chen}
\affiliation{Department of Physics, The University of Hong Kong, Pokfulam Road, Hong Kong, China}
\author{Zhizhou Yu}
\affiliation{School of Physics and Technology, Nanjing Normal University, Nanjing 210023, China}
\author{Lei Zhang}
\email[]{zhanglei@sxu.edu.cn}
\affiliation{State Key Laboratory of Quantum Optics and Quantum Optics Devices, Institute of Laser Spectroscopy, Shanxi University, Taiyuan 030006, China}
\affiliation{Collaborative Innovation Center of Extreme Optics, Shanxi University, Taiyuan 030006, China}
\author{Baigeng Wang}
\affiliation{National Laboratory of Solid State Microstructures and Department of Physics, Nanjing University, Nanjing 210093, China}
\author{Jian Wang}
\email[]{jianwang@hku.hk}
\affiliation{College of Physics and Optoelectronic Engineering, Shenzhen University, Shenzhen 518060, China}
\affiliation{Department of Physics, The University of Hong Kong, Pokfulam Road, Hong Kong, China}

\begin{abstract}
The Landau-Lifshitz-Gilbert (LLG) equation is widely used to describe magnetization dynamics. We develop a unified framework of the microscopic LLG equation based on the nonequilibrium Green's function formalism. We present a unified treatment for expressing the microscopic LLG equation in several limiting cases, including the adiabatic, inertial, and nonadiabatic limits with respect to the precession frequency for a magnetization with fixed magnitude, as well as the spatial adiabatic limit for the magnetization with slow variation in both its magnitude and direction. The coefficients of those terms in the microscopic LLG equation are explicitly expressed in terms of nonequilibrium Green's functions. As a concrete example, this microscopic theory is applied to simulate the dynamics of a magnetic Skyrmion driven by quantum parametric pumping. Our work provides a practical formalism of the microscopic LLG equation for exploring magnetization dynamics.

\end{abstract}
\maketitle

\section{Introduction}\label{sec:intr}

Single-molecule magnets (SMMs) are mesoscopic magnets with permanent magnetization, which show both classical properties and quantum properties.\cite{garg1993,friedman1996, sangregorio1997,wernsdorfer1999,wernsdorfer2005,ardavan2007,filipovic2013} SMMs are appealing due to their potential applications as memory cells and precessing units in spintronic devices.\cite{kahn1998,timm2012} Transport of SMMs coupled with leads has been investigated both experimentally\cite{heersche2006,Jomh2006,zyazin2010,roch2011} and theoretically.\cite{park2002,tretiakov2010,filipovic2013,bode2012,Fransson14,Hammar16,Nikolic2018,Nikolic2019} Transport measurements on magnetic molecules such as $\rm{Mn}_{12}$\cite{heersche2006} and $\rm{Fe}_8$\cite{Jomh2006} revealed interesting phenomena, including peaks in the differential conductance and Coulomb blockades. Dc- and ac-driven magnetization switching and noise as well as the influence on I-V characteristics were discussed in a normal metal/ferromagnet/normal metal structure.\cite{tretiakov2010} Current-induced switching of a SMM junction was theoretically studied in the adiabatic regime within the Born-Oppenheimer approximation.\cite{bode2012} It was found that magnetic exchange interactions between molecular magnets can be tuned by electric voltage or temperature bias.\cite{Fransson14} Transient spin dynamics in a SMM was investigated with generalized spin equation of motion.\cite{Hammar17} A microscopic formalism was recently proposed for consistent modeling of coupled atomic magnetization and lattice dynamics.\cite{Fransson17}

For a SMM with magnetization $\mathbf{M}$, its magnetization dynamics can be semiclassically described by the Landau-Lifshitz-Gilbert (LLG) equation of motion\cite{Gilbert04,foros2005,chudnovskiy2008,brataas2008,foros2009,swiebodzinski2010,brataas2011}
\begin{equation}
\frac{d\mathbf{m}}{dt} = -\gamma\mathbf{m}\times \mathbf{H}_{\text{eff}} + \mathbf{m}\times(\bm{\alpha}\frac{d\mathbf{m}}{dt})+ \bm{\tau}_{\rm{STT}}, \label{LLG0}
\end{equation}
where $\mathbf{m}= \mathbf{M}/M$ is the unit magnetization vector, $\gamma$ is the gyromagnetic ratio, and
$\mathbf{H}_{\text{eff}}$ is the effective magnetic field around which the magnet precesses. $\bm{\alpha}$ is the Gilbert damping tensor describing the dissipation of the precession, and $\bm{\tau}_{\rm{STT}}$ is the spin transfer torque due to the misalignment between the magnetization and the transport electron spin.\cite{slonczewski1996,berger1996,tserkovnyak2005,ralph2008} 

The LLG equation is widely adopted to describe magnetization dynamics in the adiabatic limit, where the magnetization precesses slowly and the typical time scale is in the order of $ns$. The Gilbert damping term is in general a $3 \times 3$ tensor,\cite{tserkovnyak2005} which can be deduced from experimental data, scattering matrix theory,\cite{brataas2008,brataas2011} or first-principles calculation.\cite{Starikov10,Bhattacharjee12,Mankovsky13} Later, the LLG equation was generalized to study ultrafast dynamics induced by $ps$ electrical pulse\cite{others3,Jhuria} or $fs$ laser pulse\cite{koopmans,kammerer,others1,others2}, which extends the magnetization switching time down to $ps$ or even sub-$ps$. This is refereed as the inertial regime\cite{fahnle2011}, where the time scale involved is much shorter than that of the adiabatic limit. In the inertial limit, a nonlinear inertial term was introduced into the LLG equation\cite{suhl,Ciornei,Li15,Olive15,Mondal21}, which was applied to simulate ultrafast spin dynamics.\cite{Hammar17,Mondal16,Mondal17} Direct observation of inertial spin dynamics was experimentally realized in ferromagnetic thin films in the form of magnetization nutation at a frequency of $\rm ~0.5 ~THz$.\cite{Neeraj21} When the magnetization varies in both temporal and spatial domains, two adiabatic spin torques were incorporated into the LLG equation\cite{Zhang-SF}, which can describe the dynamics of magnetic textures such as Skymions.\cite{Tserkovnyak2017}

Magnetic Skyrmions (Sk) stabilized by the Dzyaloshinskii-Moriya interaction (DMI) or competing interaction between frustrated magnets are topologically nontrivial spin textures showing chiral particle-like nature. When an electron traverses the Sk, it acquires a Berry phase and experiences a Lorentz-like force, leading to the topological Hall effect\cite{THE}. At the same time, the Magnus force due to the back action on Sk gives rise to Skyrmion Hall effect\cite{Jiang,Litzius}. The existence of Sk has been verified in magnetic materials including MnSi\cite{Muhlbauer} and PdFe/Ir(111)\cite{Wiesendanger}. The radius of an Sk can be as small as a few $nm$\cite{Heinze,Romming} and is stable even at room temperatures\cite{Moreau-Luchaire,Woo}. Sk can be operated at ultralow current density,\cite{Jonietz,Yu,YanZhou20} which makes it promising in spintronic applications including the magnetic memory and logic gates.\cite{Fert,Zhang} Various investigations show that Sk can be manipulated by spin torque due to the charge/spin current injection\cite{Jonietz,Yu}, external electric field,\cite{Upadhyaya,White} magnetic field gradient\cite{Wang2017}, temperature gradient,\cite{Kong,Mochizuki,Yanzhou20PRL,YanZhou22} and strain,\cite{Shibata} etc. However, Sk driven by quantum parametric pumping has not been explored.

Quantum parametric pump refers to such a process: in an open system without bias voltages, cyclic variation of system parameters can give rise to a net dc current per cycle\cite{brouwer,Aleiner,Altshuler,Switkes,Zhou-F,Shutenko,Wei-YD,xu2011,xu2023}. In the adiabatic limit, this quantum parametric pump requires at least two pumping parameters with a phase difference and the pumped current is proportional to the area enclosed by the trajectory of pumping parameters in parameter space.\cite{brouwer} It was found that the adiabatic pumped current is related to Berry phase.\cite{Avron} Beyond the adiabatic limit, the cyclic frequency may serve as another dimension in parameter space and hence a single-parameter quantum pump is possible at finite frequences.\cite{Vavilov,Wang-BG1} In general, quantum parametric pump can be formulated in terms of photon-assisted transport.\cite{Buttiker1,Wang-BG2} Quantum parametric pump can also generate heat current\cite{Buttiker2,Wang-BG3}, whose lower bound is Joule heating during the pumping process. This defines an optimal quantum pump\cite{Avron1,Wang-BG4} that is noiseless and pumps out quantized charge per cycle\cite{Levinson,Wang-BG5,xu2022}. Quantum parametric pumping theory has been extended to account for Andreev reflection in the presence of superconducting lead\cite{Wang-J2001}, correlated charge pump\cite{Sharma,Sun-QF1}, and parametric spin pump\cite{Wang-BG6,Zheng-W}, providing more physical insights. It is interesting to generalize quantum parametric pump to Skyrmion transport, which may offer new operating paradigms for spintronic devices.

In this work, we investigate the microscopic origin of the LLG equation and the Gilbert damping. We focus on several limiting cases of the LLG equation. For a magnetization with fixed magnitude, the adiabatic, inertial, and nonadiabatic limits with respect to its precession frequency are discussed. When both the magnitude and direction of a magnetization vary slowly in space, which is referred as the adiabatic limit in spatial domain, our formalism can also be extended to cover this limit. We will provide a unified treatment of all these cases and explicitly express each term in the microscopic LLG equation in the language of nonequilibrium Green's functions. As an example, we apply the microscopic LLG equation to simulate the dynamics of a Skyrmion driven by quantum parametric pumping in a two-dimensional (2D) system.

This paper is organized as follows. In Sec.\ref{sec:model}, a single-molecule magnet (SMM) transport setup and corresponding Hamiltonians are introduced. In Sec.\ref{sec:dynamics}, a stochastic Langevin equation for magnetization dynamics is derived from the equation of motion by separating fast (electron) and slow (magnetization) degrees of freedom, forming a microscopic version of the LLG equation. In Sec.\ref{microLLG}, four limiting cases of the microscopic LLG equation are discussed. In Sec.\ref{PumpSk}, we numerically study Sk transport driven by quantum parametric pumping. Finally, a brief summary is given in Sec.\ref{concludesec}.

\section{\label{sec:model}Model}

\begin{figure}
\centering
\includegraphics[width=8cm]{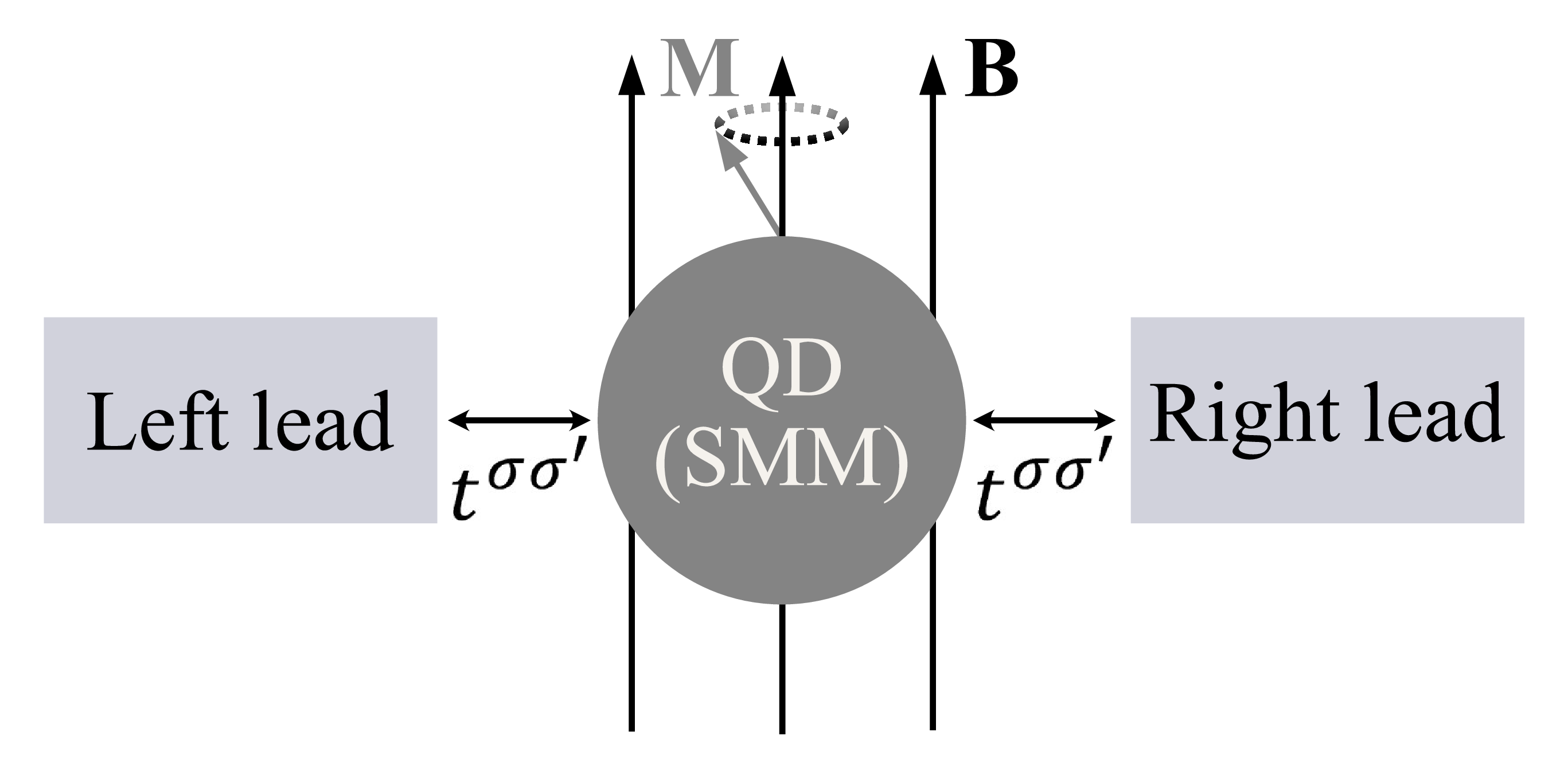}
\caption{Sketch of the model system. A single-molecule magnet (SMM) represented by the quantum dot (QD) is connected to the left and right leads. A uniform magnetic field is applied in the central region, around which the SMM magnetization precesses. }\label{fig1}
\end{figure}

The model system under investigation is shown in Fig.\ref{fig1}, where a noninteracting quantum dot (QD) representing a single-molecule magnet (SMM) with magnetization $\mathbf{M}$ is connected to two leads. A uniform magnetic field $\mathbf{B} = B \hat{e}_z$ is applied in the central region. In addition, we assume that there is a dc bias or spin bias across the system providing a spin transfer torque or spin orbit torque.

The Hamiltonian of this system is given by ($\hbar=1$)
\begin{equation}
\hat{H}_{\rm total} = \hat{H}_{L} + \hat{H}_{R} + \hat{H}_{D} + \hat{H}_{T}, \nonumber
\end{equation}
with the lead Hamiltonian ($\alpha=L,R$),
\begin{equation}
\hat{H}_{\alpha} = \sum_{k\sigma}\epsilon_{k\alpha\sigma} \hat{c}^{\dagger}_{k\alpha\sigma} \hat{c}_{k\alpha\sigma}, \label{lead}
\end{equation}
and the Hamiltonian of the central region,
\begin{equation}
\hat{H}_D = \hat{H}_0 + \hat{H}' + \gamma\hat{\mathbf{M}}\cdot\mathbf{B}. \label{eqHD}
\end{equation}
Here $\hat{H}_0$ is the Hamiltonian of the QD with spin-orbit interaction (SOI)\cite{sun2005}
\begin{equation}
\hat{H}_0=\sum_{n\sigma}\epsilon_{n\sigma}\hat{d}^{\dagger}_{n\sigma}\hat{d}_{n\sigma}+\sum_{mn}(t^{SO}_{nm} d^\dagger_{m\uparrow} d_{n\downarrow}+\rm{H.c.}),\label{H0}
\end{equation}
with $t^{SO}_{nm}= -t^{SO}_{mn}$. $\hat{H}'$ is the interaction between the electron spin and the magnetization as well as the magnetic field,
\begin{equation}
\hat{H}'= J \sum_n\hat{\mathbf{s}}_n\cdot\hat{\mathbf{M}} +\gamma_e\sum_{n} \hat{\mathbf{s}}_n\cdot\mathbf{B}.\nonumber
\end{equation}
We can also add uniaxial anisotropy field to $\hat{H}'$. The coupling Hamiltonian between the QD and the leads is
\begin{equation}
\hat{H}_{T}=\sum_{k\alpha n,\sigma\sigma'}[t_{k\alpha n}^{\sigma\sigma'}\hat{c}^{\dagger}_{k\alpha\sigma} \hat{d}_{n\sigma'}+\rm{H.c.}].\label{eqHT}
\end{equation}

In the above equations, $\hat{d}^{\dagger}_{n\sigma}$ ($\hat{c}^{\dagger}_{k\alpha\sigma}$) creates an electron with energy $\epsilon_{n\sigma}$ ($\epsilon_{k\alpha\sigma}$) in the QD (lead $\alpha$). In general, the leads can be metallic or ferromagnetic. Here $\hat{\mathbf{s}}_{n} = \frac{1}{2}\psi_{n}^{\dagger}\bm{\sigma}\psi_{n}$ is the electron spin in the central region, with $\psi^\dagger_{n} = (d^\dagger_{n\uparrow},d^\dagger_{n\downarrow})$. The Pauli matrices satisfy $[\sigma_x,\sigma_y] = 2i\sigma_z$, and the magnetization $\mathbf{M}$ follows the commutation relation $[M_x,M_y] = i\hbar M_z$. $J$ is the exchange interaction between the magnetization and the spin of conducting electrons. $\gamma$ ($\gamma_e$) is the gyromagnetic ratio of the magnet (electron).

If we choose the magnetic field in the $z$ direction as the laboratory frame, and $(\theta,\phi)$ the polar and azimuthal angles of the magnetization, the spin dependent coupling matrix is given by
\begin{equation}
t_{k\alpha n}^{\sigma\sigma'} = \big[ \hat{R} t_{k\alpha n} \big]_{\sigma\sigma'},\label{tunnel01}
\end{equation}
with $\hat{R}$ the rotational operator\cite{kamenev2011field}
\begin{equation}
\hat{R} =
e^{-i\frac{\theta}{2}\hat{\sigma}_y} e^{-i\frac{\phi}{2}\hat{\sigma}_z} =
\begin{pmatrix}
      e^{-i\frac{\phi}{2}}\cos(\frac{\theta}{2}) &
      -e^{i\frac{\phi}{2}}\sin(\frac{\theta}{2}) \\
      e^{-i\frac{\phi}{2}}\sin(\frac{\theta}{2}) &
      e^{i\frac{\phi}{2}}\cos(\frac{\theta}{2})
\end{pmatrix}. \label{tunnel02}
\end{equation}

\section{\label{sec:dynamics}magnetization dynamics}

From the Heisenberg equation of motion, the magnetization dynamics in the central region is governed by
\begin{equation}
\dot {\hat{\mathbf{M}}} = -\gamma\hat{\mathbf{M}}\times\mathbf{B} - J\hat{\mathbf{M}}\times\hat{\mathbf{s}}_D, \label{dynamic01}
\end{equation}
where $\hat{\mathbf{s}}_D = \sum_{n}\hat{\mathbf{s}}_n$ is the total electron spin. In deriving the above equation, the following relation is used:
\begin{equation}
[\hat{\bm{\sigma}},\hat{\bm{\sigma}}\cdot\mathbf{A}] = -2i\hat{\bm{\sigma}}\times\mathbf{A}. \label{sigmaRelation01}
\end{equation}

Now we separate an operator into its quantum average and its fluctuation, then $\hat{\mathbf{s}}_D = \langle\hat{\mathbf{s}}_D\rangle + \delta \hat{\mathbf{s}}_D$, and $\hat{\mathbf{M}} = \langle\hat{\mathbf{M}}\rangle + \delta \hat{\mathbf{M}}$, where $\delta \hat{\mathbf{s}}_D$ ($\delta \hat{\mathbf{M}}$) is the fluctuation of the electron (magnet) spin. We can transform Eq.~(\ref{dynamic01}) into a Langevin equation. For the expectation value $\mathbf{M}(t) = \langle\hat{\mathbf{M}}(t)\rangle$,\cite{timescale}
\begin{equation}
\dot{\mathbf{M}} = \mathbf{M}\times[-\gamma \mathbf{B} - J\mathbf{s}_D + \delta \mathcal{{\hat B}} ], \label{dynamic02}
\end{equation}
or
\begin{equation}
\dot{\mathbf{M}} = -\gamma\mathbf{M}\times[\mathbf{H}_{\text{eff}} - \delta \mathcal{{\hat B}}^{'} ], \nonumber
\end{equation}
where
\begin{equation}
\mathbf{s}_D = \langle\hat{\mathbf{s}}_D\rangle = -\frac{i}{2}\mathrm{Tr}[\bm{\sigma}G^{<}(t,t)]. \label{sigma-dot}
\end{equation}
Here $G^<_{ij\sigma\sigma'}(t',t) = i\langle d^\dagger_{j\sigma'}(t') d_{i\sigma}(t) \rangle$ is the lesser Green's function of electrons, which will be discussed in detail below. The effective magnetic field $\mathbf{H}_{\text{eff}}$ is defined as the variation of the free energy of the system with respect to the magnetization\cite{berkov2007,gilmore2008,tserkovnyak2005}
\begin{equation}
\mathbf{H}_{\text{eff}} = \frac{1}{\gamma}\frac{\delta H_{\rm total}}{\delta \mathbf{M}}. \label{eqHeff}
\end{equation}
And $\delta\mathcal{{\hat B}} = \gamma \delta\mathcal{{\hat B}}^{'}$ contributes from the fluctuations
\begin{eqnarray}
{\mathbf{M}} \times \delta\mathcal{{\hat B}}&=& -\delta\dot{\hat{\mathbf{M}}} -\gamma\delta\hat{\mathbf{M}} \times {\bf{B}} - J {\mathbf{M}} \times \delta\hat{\mathbf{s}}_D \nonumber \\
& & - J\delta\hat{\mathbf{M}}\times {\mathbf{s}}_D -J\delta\hat{\mathbf{M}}\times \delta\hat{\mathbf{s}}_D. \nonumber
\end{eqnarray}
These fluctuations can play an important role in determining the motion of the magnetization, such as reducing or enhancing the threshold bias of magnetization switching.\cite{bode2012}

To transform Eq.~(\ref{dynamic02}) into the usual LLG equation, we further separate $\mathbf{s}_D$ in Eq.~(\ref{sigma-dot}) into the time-reversal symmetric and antisymmetric components, $\mathbf{s}^s_D$ and $\mathbf{s}^a_D$. Then $\mathbf{M}\times\mathbf{s}^s_D$ and $\mathbf{M}\times\mathbf{s}^a_D$ correspond to the dissipative and dissipativeless terms, respectively. Thus, Eq.~(\ref{dynamic02}) is rewritten as
\begin{equation}
\dot{\mathbf{M}} = -\gamma \mathbf{M}\times \mathbf{B} - J\mathbf{M}\times\mathbf{s}^a_D - J\mathbf{M}\times\mathbf{s}^s_D. \label{dynamic03}
\end{equation}
Note that the last term in Eq.~(\ref{dynamic03}), ${\bf M} \times \mathbf{s}_D^{s}$, corresponds to the damping of magnetization. As will be discussed below that in the adiabatic approximation, it assumes the form $\mathbf{M}\times ({\bm \alpha}\dot{\mathbf{M}})$ where $\bm \alpha$ is the Gilbert damping tensor which is expressed in terms of nonequilibrium Green's function (see Eq.~(\ref{gilbert01})). The second term in Eq.~(\ref{dynamic03}), ${\bf M} \times \mathbf{s}_D^{a}$, corresponds to the spin transfer torque. In the presence of SOI, ${\bf M} \times \mathbf{s}_D^{a}$ is the spin orbit torque in collinear ferromagnetic systems, which has field-like and damping-like components, respectively, along the directions ${\bf M} \times {\bf u}$ and ${\bf M} \times ({\bf M} \times {\bf u})$ with ${\bf u} \cdot {\bf M}=0$. Here ${\bf u}$ is the unit vector of the spin current.\cite{slonczewski1996}

\section{Microscopic LLG equation in different limits }\label{microLLG}

In this section, we will drive the LLG equation and express the Gilbert damping tensor in terms of the nonequilibrium Green's functions. We also discuss the fluctuation in the equation of motion and the spin continuity equation, showing that the spin transfer torque is insufficient to describe magnetization dynamics in general conditions.

We focus on several limiting cases of the microscopic LLG equation (Eq.~(\ref{dynamic03})). These cases correspond to different limits: (1) Adiabatic limit in temporal domain where the precessing frequency of the magnet is low and $\mathbf{s}_D$ can be expanded up to the first order in frequency; (2) Inertial regime where the time scale is much shorter than that of the adiabatic limit, e.g., magnetization switching in $ps$ or even sub-$ps$ range\cite{others3,Jhuria,koopmans,kammerer,others1,others2}; (3) Nonadiabatic regime where adiabatic approximation in temporal domain is removed. We will work on the linear coupling between the magnetization and the environment,\cite{Sayad15} and derive the Gilbert damping coefficient as a function of the precessing frequency; (4) In the above situations, we have assumed that the magnetization has fixed magnitude and only its direction varies in space. Our theory can be easily extended to address the motion of domain walls where the magnetization is nonuniform. In the simplest case, we assume that the magnetization varies slowly in space so that adiabatic approximation in spatial domain can be taken. In this spatial adiabatic limit, two additional toques are incorporated into the LLG equation which are naturally obtained in our theory.

\subsection{ Adiabatic limit }\label{subsec:low_frequency}
As the magnetization precesses, the electron spin and hence spin-orbit energy of each state changes \cite{tserkovnyak2005,gilmore2008}, which drives the system out of equilibrium. In the language of frozen Green's functions (Eqs.~(\ref{lesser}) and (\ref{GF4})), total spin of the QD (Eqs.~(\ref{sigma-dot})) can be expanded in terms of the precession frequency, which consists of two parts: the quasi-static part ${\bf s}_D^{(0)}$, and the adiabatic change ${\bf s}_D^{(1)}$ to the first order in frequency
\begin{equation}
{\bf s}_D = {\bf s}_D^{(0)} + {\bf s}_D^{(1)}, \nonumber
\end{equation}
where
\begin{equation}
{\bf s}_D^{(0)} = -\frac{i}{2}\int\frac{dE}{2\pi} \mathrm{Tr}[{\bm \sigma}G_f^{<}], \label{eq_sD0}
\end{equation}
and
\begin{equation}
{\bf s}_D^{(1)} = -\frac{1}{4}\int\frac{dE}{2\pi}\mathrm{Tr}[G_f^{<}\bm{\sigma} G_f^r G_f^r\bm{\sigma}- G_f^a G_f^a\bm{\sigma}G_f^{<}\bm{\sigma}] \dot{\bf{b}},\label{eq_sD1}
\end{equation}
with $\dot{\mathbf{b}} = JM\dot{\mathbf{m}}$.

Concerning the magnetization dynamics, the effective field $\mathbf{H}_{\text{eff}}(t) $ (Eq.~(\ref{eqHeff})) can be separated into two contributions: an anisotropy field and a damping field \cite{gilmore2008},
\begin{equation}
\mathbf{H}_{\text{eff}}(t) =\mathbf{H}_{\text{eff}}^{\text{ani}}(t) +\mathbf{H}_{\text{eff}}^{\text{damp}}(t), \label{eqHeff02}
\end{equation}
with
\begin{equation}
\mathbf{H}_{\text{eff}}^{\text{ani}}(t)=\mathbf{B}+(J/\gamma)\mathbf{s}_D^{(0)}, ~ ~~~\mathbf{H}_{\text{eff}}^{\text{damp}}(t)=(J/\gamma)\mathbf{s}_D^{(1)}.\label{eqHeff02-0}
\end{equation}

Substituting Eqs.~(\ref{eq_sD0}) and (\ref{eq_sD1}) into Eq.~(\ref{dynamic03}), ignoring the fluctuation, and noting that $\dot{\mathbf{b}} = JM\dot{\mathbf{m}}$, we obtain the deterministic Landau-Lifshitz-Gilbert equation,
\begin{equation}
\frac{d\mathbf{m}}{dt} = -\gamma\mathbf{m}\times \mathbf{H}^{\text{ani}}_{\text{eff}} - \mathbf{m}\times({\bm \alpha}\dot{\mathbf{m}}), \label{LLG01}
\end{equation}
where
\begin{equation}
\mathbf{m} = \mathbf{M}/M = \big(\sin\theta\cos\phi,\sin\theta\sin\phi,\cos\theta\big),\nonumber
\end{equation}
is the unit vector in the magnetization direction. $\bf{\alpha}$ is the $3\times 3$ Gilbert damping tensor\cite{fahnle2011,brataas2008}, which is defined in terms of the frozen Green's functions:
\begin{equation}
\alpha_{ij} = \frac{(JM)^2}{4} \int\frac{dE}{2\pi}\mathrm{Re}\{\mathrm{Tr}[G_f^{<} \sigma_i G_f^r G_f^r \sigma_j ]\}. \label{gilbert01}
\end{equation}
As shown in Appendix D, this damping tensor recovers that obtained in Ref.~[\onlinecite{brataas2008}] via the scattering matrix theory in the limit of zero temperature and in the absence of external bias. In general, the Gilbert damping tensor depends on ${\bf m(t)}$ and bias voltage through the frozen Green's functions $G^r_f$ and $G^<_f$. This agrees with the observation in Ref.~[\onlinecite{steiauf}] using the effective field theory of breathing Fermi surface mode.

\subsection{Inertial regime}

In this regime, the magnetization has both precessional and nutational motions. We focus on the linear coupling between the magnetization and the environment so that an additional ``inertial" term enters the LLG equation, which describes the nutation of the magnet. In this case, the adiabatic approximation is not good enough. One has to expand the spin density ${\bf s}_D$ at least to the second order in frequency. In the inertial regime, we assume that the magnitude of the magnetization is fixed while only its direction varies. Iterating Eq.~(\ref{GF4}) to the second order in frequency, we have
\begin{equation}
G^r = G^r_f -i G^r_f {\dot G}^r_f + G^r_f ({\dot G}^r_f)^2 + (G^r_f)^2 {\ddot G}^r_f, \nonumber
\end{equation}
from which we find the contribution ${\bf s}^{(2)}_D$ in the inertial limit,
\begin{eqnarray}
{\bf s}_D^{(2)} &=& -\frac{1}{8}\int\frac{dE}{2\pi}{\rm Im} \{\mathrm{Tr}[G_f^{<}\bm{\sigma}
(G_f^r)^3 \bm{\sigma} ]\} \cdot \partial^2_t {\bf{b}} \nonumber \\
&+& \frac{1}{16}\int\frac{dE}{2\pi}{\rm Im} \{\mathrm{Tr}[G_f^{<}\bm{\sigma}
(G_f^r)^4 (\bm{\sigma} \cdot \partial_t {\bf{b}})^2] \}\nonumber \\
&-& \frac{i}{8}\int\frac{dE}{2\pi}\mathrm{Tr}[\bm{\sigma} (G_f^r)^2 G^<_f  (G_f^a)^2  (\bm{\sigma} \cdot \partial_t {\bf{b}})^2]. \label{eq_sD2}
\end{eqnarray}
To make comparison with Ref.~[\onlinecite{fahnle2011}], we keep only the linear term $\partial^2_t {\bf m}$ and neglect other nonlinear terms such as $(\partial_t {\bf m})^2$. With this new term, the LLG equation in inertial regime is written as\cite{suhl,fahnle2011,Li15,Ciornei,Tserkovnyak2017}
\begin{equation}
\frac{d\mathbf{m}}{dt} = -\gamma\mathbf{m}\times \mathbf{H}^{\text{ani}}_{\text{eff}} -
\mathbf{m}\times[\bm{\alpha}\frac{d\mathbf{m}}{dt}]- \mathbf{m}\times[\bm{{\bar \alpha}}\frac{d^2\mathbf{m}}{dt^2}], \label{LLG02}
\end{equation}
where the inertial term is a $3\times 3$ tensor given by
\begin{equation}
{\bar \alpha}_{ij} = \frac{(JM)^2}{2} \int\frac{dE}{2\pi}\mathrm{Im}\{\mathrm{Tr} [G_f^{<} \sigma_i (G_f^r)^3 \sigma_j ]\}. \label{gilbert02}
\end{equation}
This additional inertial term has been obtained both phenomenologically\cite{suhl} and semiclassically\cite{fahnle2011}. Ref.~[\onlinecite{Bhattacharjee12}] proposed a first-principles method for calculating the inertia term in the semiclassical limit. Here we derive the quantum inertial tensor in terms of the frozen Green's function.


\subsection{\label{subsec:weak_magnetization} Nonadiabatic regime}
Now we consider the magnetization dynamics at finite precession frequency, whose time scale is still much larger than that of electrons. Since no analytic solution exists in general conditions, we only focus on the linear coupling in the exchange interaction $J$. In this nonadiabatic regime, we treat the coupling $J$ as a small perturbation, and rewrite the equation determining the Green's function as
\begin{equation}
\big( i\frac{\partial}{\partial t} - \tilde{H}_0 - H'(t)-\Sigma^r \big) G^r(t,t') = \delta(t-t'), \label{dyson02}
\end{equation}
where $\tilde{H}_0 = H_0 + \gamma_e {\bf s}_D \cdot\mathbf{B}$ is the unperturbed Hamiltonian, including the bare Hamiltonian of the QD defined in Eq.~(\ref{H0}) and the Hamiltonian due to the constant external field. $H'(t)=J \bm{\sigma} \cdot\mathbf{M}(t)$ is the perturbative term due to exchange coupling between the magnetization and the electron spin.

The unperturbed retarded Green's function satisfies
\begin{equation}
(i\frac{\partial}{\partial t} - \tilde{H}_0 -\Sigma^r ) G^r_0(t-t') = \delta(t-t'). \label{dyson03}
\end{equation}

Since $G^r_0(t-t')$ only depends on the time difference, it is convenient to work in the energy representation,
\begin{equation}
G^r_0(E) = [E - H_0 -\frac{\gamma_e}{2} \bm{\sigma}\cdot\mathbf{B} - \Sigma^r]^{-1}, \label{Gr0}
\end{equation}
where $G^r_0(t-t')$ and $G^r(t,t')$ are related through the Dyson equation
\begin{equation}
G^r(t,t') = G^r_0(t-t') + \int dt_1 G^r_0(t-t_1) H'(t_1)G^r(t_1,t'). \nonumber
\end{equation}

In the first order perturbation, we have
\begin{equation}
G^r = G^r_0 + G^r_0 H' G^r_0, \nonumber
\end{equation}
and
\begin{equation}
G^{<} = G^{<}_0 + G^r_0 H' G^{<}_0 + G^{<}_0 H' G^a_0. \nonumber
\end{equation}

Using
\begin{equation}
\begin{split}
&\ \ \ \
\int dt_1 G^r_0(t-t_1)
H'(t_1)
G^{<}_0(t_1-t) \\ &  =
\int\frac{dE}{2\pi}\int\frac{d\omega}{2\pi}
e^{-i\omega t}
G^r_0(E+\omega)H'(\omega)G^{<}_0(E),\nonumber
\end{split}
\end{equation}
and
\begin{equation}
\begin{split}
&\ \ \ \
\int dt_1 G^{<}_0(t-t_1)
H'(t_1)
G^a_0(t_1-t) \\ &  =
\int\frac{dE}{2\pi}\int\frac{d\omega}{2\pi}
e^{-i\omega t}
G^{<}_0(E+\omega)H'(\omega)G^{a}_0(E),\nonumber
\end{split}
\end{equation}
where $H'(\omega) = \frac{J}{2}\bm{\sigma} \cdot \mathbf{M}(\omega)$ with $\omega$ the precession frequency, the spin density $\mathbf{s}_D$ of the quantum dot can be evaluated
\begin{equation}
{\bf s}_D = {\bf s}_D^{(0)} + {\bf s}_D^{(1)}.
\end{equation}
Here ${\bf s}_D^{(0)}$ is independent of $\mathbf{M}$ and time $t$:
\begin{equation}
{\bf s}_D^{(0)} = -\frac{i}{2}\int\frac{dE}{2\pi} \mathrm{Tr}[\bm{\sigma}G_0^{<}(E)].\nonumber
\end{equation}
And ${\bf s}_D^{(1)}$ depends linearly on $\mathbf{M}(t)$
\begin{equation}
\begin{split}
{\bf s}_D^{(1)} = &
-\frac{i}{2}
\int\frac{dE}{2\pi}\int\frac{d\omega}{2\pi}
e^{-i\omega t} \mathrm{Tr}[
\bm{\sigma}G^r_0(E+\omega)H'(\omega)G^{<}_0(E) \\ &
+ \bm{\sigma}G^{<}_0(E+\omega)H'(\omega)G^{a}_0(E) ].\nonumber
\end{split}
\end{equation}
Using the anisotropic field $\mathbf{H}_{\text{eff}}^{\text{ani}}(t)$ and the damping field $\mathbf{H}_{\text{eff}}^{\text{damp}}(t)$ expressed in Eq.~(\ref{eqHeff02-0}) and ignoring the fluctuations, we can obtain a deterministic dynamic equation from Eq.~(\ref{dynamic03}):
\begin{equation}
\frac{d\mathbf{m}}{dt}=-\gamma\mathbf{m}\times \mathbf{H}^{\text{ani}}_{\text{eff}} - \gamma\mathbf{m}\times \mathbf{H}_{\text{eff}}^{\text{damp}},
\end{equation}
where
\begin{equation}
\mathbf{H}_{\text{eff}}^{\text{ani}}(t)= \mathbf{B} -\frac{iJ}{2\gamma}\int\frac{dE}{2\pi}\mathrm{Tr}[\bm{\sigma}G_0^{<}(E)],
\end{equation}
and
\begin{equation}
\mathbf{H}_{\text{eff}}^{\text{damp}} (t) =- \int\frac{d\omega}{2\pi} e^{-i\omega t} \mathbf{m}({\omega}) {\tilde \alpha}(\omega).
\end{equation}
Here ${\tilde \alpha}(\omega) $ is the frequency dependent Gilbert damping tensor defined as
\begin{eqnarray}
{\tilde \alpha}(\omega)&=&\frac{i}{4}J^2 M^2\int\frac{dE}{2\pi} \mathrm{Tr} \biggl[ G^{<}_0(E) \bm{\sigma}G^r_0(E+\omega)\bm{\sigma} \nonumber \\
&+& G^{a}_0(E)\bm{\sigma}G^{<}_0(E+\omega)\bm{\sigma} \biggr].
\end{eqnarray}
It is easy to confirm that when $\omega$ goes to zero, we can recover the results in the adiabatic and inertial limits.


\subsection{Adiabatic limit in spatial domain}

When both the magnitude and direction of the magnetization vary slowly in space, we refer to this situation as the adiabatic limit in spatial domain. In this case, two additional terms emerge in the LLG equation\cite{Zhang-SF},
\begin{eqnarray}
\frac{d\mathbf{m}}{dt} &=& -\gamma\mathbf{m}\times \mathbf{H}^{\text{ani}}_{\text{eff}} -
\mathbf{m}\times[\bm{\alpha}\frac{d\mathbf{m}}{dt}]\nonumber \\
&+& b_J ({\bm j}_e \cdot \nabla)\mathbf{m} -c_J \mathbf{m}\times({\bm j}_e \cdot \nabla)\mathbf{m}, \label{LLG03}
\end{eqnarray}
where $b_J$ and $c_J$ are constants defined in Ref.~[\onlinecite{Zhang-SF}]. Here the term with coefficient $b_J$ is related to the adiabatic process of the nonequilibrium conducting electrons.\cite{Zhang-SF} In contrast, the other term with coefficient $c_J$ corresponds to the nonadiabatic process which changes sign upon time-reversal operation.

In this limit, the coupling between the magnetization and the electron spin can be approximated as
\begin{equation}
\hat{H}'= J \hat{\mathbf{s}}_{\bf r}\cdot \hat{\mathbf{M}}({\bf r},t) +\gamma_e \hat{\mathbf{s}}_{\bf r}\cdot\mathbf{B}. \label{local}
\end{equation}
Eqs.~(\ref{dynamic03}), (\ref{continuity01}), and (\ref{dynamic04}) are still valid except that ${\bf M}$ and ${\bm s}_D$ are local variables depending on position, where ${\bm s}_D(x)$ is defined as
\begin{equation}
{\bf s}_D(x) = -\frac{i}{2}\int\frac{dE}{2\pi} \mathrm{Tr}_s[{\bm \sigma}G^{<}]_{xx}, \label{sDx}
\end{equation}
where the trace is taken only in spin space.

To derive the adiabatic term in Eq.~(\ref{LLG03}), we start from Eq.~(\ref{continuity01}) and then use Eq.~(\ref{dynamic04}). From Eq.~(\ref{continuity01}), we have\cite{note9}
\begin{equation}
\frac{d{\mathbf{s}}_D}{dt} + \nabla \cdot {\bm j}_s = J{\mathbf{M}}\times {\mathbf{s}}_D,
\label{continuity02}
\end{equation}
where ${\bm j}_s$ is the spin current density and the term $-\gamma_e{\mathbf{s}}_D\times\mathbf{B}$ is neglected. Using the fact that ${\bm j}_s \approx -b_0 {\bm j}_e {\bm m}$ (where $b_0 = \mu_B P/e$ and $P$ is the polarization) and neglecting the second order terms such as $\partial_t \delta {\mathbf{s}}_D$, we find from Eq.~(\ref{continuity02})\cite{note10}
\begin{equation}
J{\mathbf{M}} \times\delta{\mathbf{s}}_D =  -b_0({\bm j}_e \cdot \nabla){\bm m},
\end{equation}
where $\delta{\mathbf{s}}_D$ denotes the contribution due to the spatial variation of the magnetization $\nabla {\bm m}$. The nonadiabatic term can be generated by iterating the following equation,
\begin{equation}
\frac{d\mathbf{m}}{dt} = -\gamma\mathbf{m}\times
\mathbf{H}^{\text{ani}}_{\text{eff}} - \mathbf{m}\times[\bm{\alpha}\frac{d\mathbf{m}}{dt}]+ b_J ({\bm j}_e \cdot \nabla)\mathbf{m}, \label{LLG04}
\end{equation}
from which we arrive at\cite{berkov2007},
\begin{eqnarray}
\frac{d\mathbf{m}}{dt} &=&-\frac{\gamma}{1+\alpha^2} \mathbf{m}\times \mathbf{H}^{\text{ani}}_{\text{eff}} -\frac{\gamma\alpha}{1+\alpha^2}
\mathbf{m}\times[\mathbf{m}\times\mathbf{H}^{\text{ani}}_{\text{eff}}] \nonumber \\
&+& \frac{b_J}{1+\alpha^2} ({\bm j}_e \cdot \nabla)\mathbf{m} +\frac{b_J\alpha}{1+\alpha^2} \mathbf{m}\times({\bm j}_e \cdot \nabla)\mathbf{m},
\end{eqnarray}
where we have assumed that the Gilbert damping tensor $\bm{\alpha}$ is diagonal, i.e., $\alpha_{ij}=\alpha\delta_{ij}$. The nonadiabatic term can also be derived explicitly, as shown in Appendix E.


\section{\label{PumpSk} Skyrmion dynamics driven by quantum parametric pumping}

\begin{figure}
\centering
\includegraphics[width=8cm]{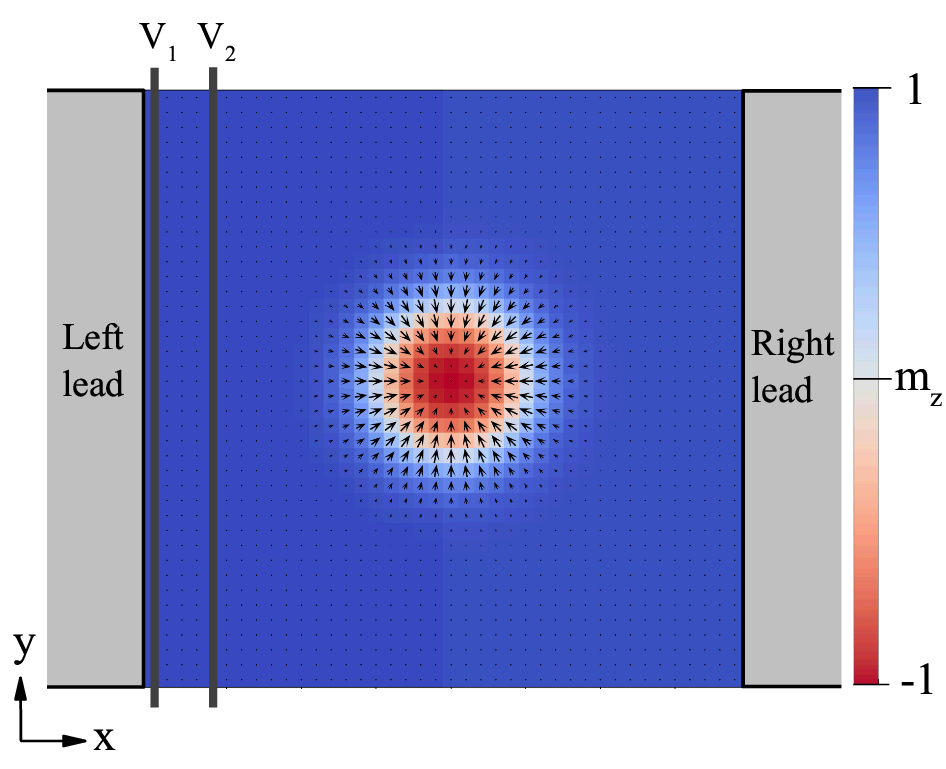}%
\caption{Schematic plot of a central region that hosts a Skyrmion and is connected to two metallic leads. The central region consists of a square lattice of size 40$\times$40. The arrows denote the in-plane component of the magnetization texture of the Skyrmion. Pumping potentials $V_1$ and $V_2$ are applied on the first and fifth column layers of the central region, which are labeled by dark gray bars.}\label{setup}
\end{figure}

In this section, we apply our microscopic theory to investigate Skyrmion dynamics in a 2D system driven by quantum parametric pumping. Initially, an Sk is placed in the central region of a two-lead system, as shown in Fig.~\ref{setup}. Then, we apply two time-dependent voltage gates with a phase difference in the system to drive a dc electric current. The electron flow, in turn, interacts with the Sk, which gives rise to quantum parametric pumping of the Sk. In the tight-binding representation, the Sk is described by the following Hamiltonian
\begin{eqnarray}
H_{Sk} = &-&J_{ex} \sum_{\langle i,j\rangle} \mathbf{m}_i\cdot \mathbf{m}_j + \sum_{\langle i,j\rangle} \mathbf{D} \cdot(\mathbf{m}_i \times \mathbf{m}_j) \nonumber \\
         &-& K \sum_i (\mathbf{m}_i\cdot {\hat z})^2- \mu \sum_i\mathbf{m}_i\cdot \mathbf{B}. \label{SkHam}
\end{eqnarray}
Here $J_{ex}$ is the Heisenberg exchange interaction. $\mathbf{D}=D ({\bm r}_i -{\bm r}_j)/|{\bm r}_i -{\bm r}_j|$ is the Dzyaloshinskii-Moriya interaction (DMI). $K$ is the perpendicular magnetic anisotropy constant, and $\mu$ is the magnitude of the magnetic moment. To facilitate parametric pumping, we apply gate voltages in two different regions of the system with the following form,
\begin{equation}
V_{p} = V_1 \cos(\omega_p t) + V_2 \cos(\omega_p t+\phi), \nonumber
\end{equation}
where $V_1= V \delta(x-l_1)$ and $V_2=V \delta(x-l_2)$ are potential landscapes with $V$ the pumping amplitude, $\omega_p$ is the pumping frequency, and $\phi$ is the phase difference. The central scattering region is discretized into a $40 \times 40$ mesh. The positions of gate voltages are $l_1=1$ and $l_2=5$, which are displayed in Fig.~\ref{setup}. In the adiabatic pumping regime (small $\omega_p$ limit), the cyclic variation of two potentials $V_1$ and $V_2$ can pump out a net current when $\phi \neq n\pi$\cite{brouwer,Wei-YD}. Thus, the total Hamiltonian of the system consists of $H_\alpha$, $H_0$, $H_T$, $H'$ (given by Eqs.~(\ref{lead}), (\ref{H0}), (\ref{eqHT}), and (\ref{local}), respectively), $H_{Sk}$, and $V_p$.

Since the Sk has slow varying spin texture in space, its dynamics can be approximated by the adiabatic limit in spatial domain. The following LLG equation describing the Sk dynamics driven by parametric pumping needs to be solved,
\begin{equation}
\frac{d\mathbf{m}_i}{dt} = -\gamma\mathbf{m}_i\times [\mathbf{H}_{\rm eff}-(J/\gamma)\mathbf{s}_D] -
\mathbf{m}_i\times({\bm \alpha}\dot{\mathbf{m}_i}), \label{LLG07}
\end{equation}
where the effective field $\mathbf{H}_{\text{eff}}$ is defined as
\begin{equation}
\mathbf{H}_{\text{eff}} =\frac{1}{\gamma}\frac{\delta H_{Sk}}{\delta \mathbf{m}_i}. \nonumber
\end{equation}
Here $\mathbf{s}_D$ is defined in terms of Green's functions in Eq.~(\ref{sDx}). The Gilbert damping tensor ${\bm \alpha}$ is assumed to be a diagonal matrix, $\alpha_{ij}=\alpha\delta_{ij}$. It is worth mentioning that Eq.~(\ref{LLG07}) already includes the $({\bm j}_e \cdot \nabla) \mathbf{m}$ and  $\mathbf{m} \times ({\bm j}_e \cdot \nabla)\mathbf{m}$ terms, which is discussed in Sec.~{\color{blue}IV D} and Appendix E.


Initial configuration of the Sk is generated by manually creating a topological unity charge at the center of the system and then relaxing the spin texture numerically until the magnetic energy is stable. Note that $m_z$ at the Sk center is negative, while the outside values are positive. In numerical simulation, the central region is a $40a\times 40a$ square lattice with $a$ the lattice spacing; the relaxed Sk radius is $r_0=10$, which is the minimal distance between the Sk center $m_i^z(0)=-1$ and $m_i^z(r_0)=1$. Parameters are set as $D=0.2 J_{ex}$\cite{foros2009}, $K=0.07 J_{ex}$\cite{foros2009}, $J=2 J_{ex}$, $B=0$, and $\alpha=0.4$. The Heisenberg exchange constant $J_{ex}=t=1$ is chosen as the energy unit, where $t$ is the hopping energy. We set $\hbar=\gamma=a=1$, and then the coefficients to convert the time $t$, current density $\kappa$, and velocity $v$ to SI units are $\hbar/J_{ex}$, $2eJ_{ex}/(a^2\hbar)$, and $J_{ex}a/h$.\cite{Wangw2015,Zhangb2015} Table \ref{tab1} shows the expressions and particular values for $J_{ex}=1$ meV and $a=0.5$ nm.

\begin{table}
\caption{\label{tab1} Unit conversion table for $J_{ex}=1$ meV and $a=0.5$ nm.}
\begin{ruledtabular}
\begin{tabular}{lll}
Distance $x$ & $\hat{x} = a$ & $ = 0.5$ nm\\
Time $t$ & $\hat{t} = \hbar/J_{ex}$ & $\approx 0.66$  ps \\
Current density $\kappa$ & $\hat{\kappa} = 2eJ_{ex}/a^{2}\hbar $ & $\approx 2\times 10^{12}$  A/m$^{2}$ \\
Velocity $v$ & $\hat{v} = J_{ex}a/(h)$ & $\approx 7.59 \times 10^{2}$ m/s\\
\end{tabular}
\end{ruledtabular}
\end{table}

\begin{figure}[tbp]
\centering
\includegraphics[width=8.5cm]{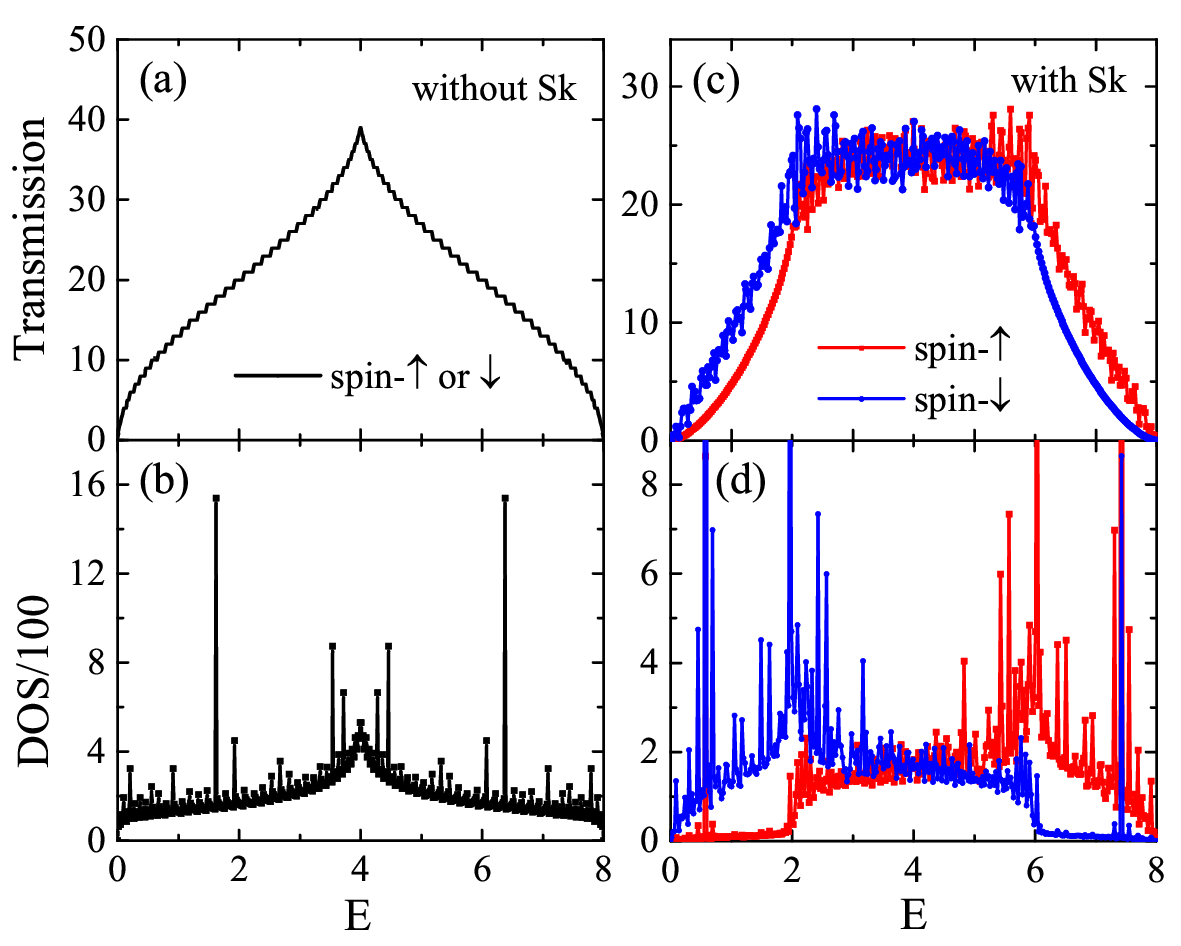}%
\caption{The transmission coefficient (a) and density of states (b) as a function of the electron energy in the absence of an Sk. (c) and (d) are the transmission and density of states for cases with an Sk at the center. No pumping potential is added in the system ($V=0$).}\label{TDOS}
\end{figure}

Our numerical calculation proceeds as follows. First, with the initial Sk configuration chosen at $t=t_0$, we calculate the total Hamiltonian of the system and then the frozen Green's function in Eq.~(\ref{GF6}) that determines $\mathbf{s}_D$ in Eq.~(\ref{sDx}). Second, the LLG equation in Eq.~(\ref{LLG07}) is solved by using the fourth-order Runge-Kutta method with a small time step $dt$. Then the Sk Hamiltonian in Eq. (\ref{SkHam}) can be updated. We repeat the above two-step calculation to simulate the Sk dynamics driven by quantum parametric pumping, and monitor the pumped current during the time evolution.

First, we investigate the static transport properties of the system without pumping. Fig.~\ref{TDOS} shows the transmission coefficient and density of states (DOS) as a function of the electron energy $E$ with and without an Sk locating at the system center. When there is no Sk, Fig.~\ref{TDOS}(a) and (b) show spin-degenerate transmission coefficients and DOS, which are standard transport properties for a metallic square lattice. However, in the presence of the Sk, spin degeneracy of the system is lifted. In Fig.~\ref{TDOS}(d), the whole energy range $[0, 8]$ can be typically divided into the following three regions, irrespective to the exchange strength $J$\cite{Ndiaye,Yin}.

(i) $0<E<|J|$. The conduction electrons are fully spin-polarized. Since J =2 in our calculation, this region corresponds to $0<E<2$. Only spin-down electrons can transmit in this energy region, and the largest spin polarization is reached near $E=2$.

(ii) $|J| <E < 8-|J|$. Both spin-up and spin-down conduction electrons exist in the system.

(iii) $8-|J|< E <8$. The conduction electrons are fully polarized with spin up component.

\begin{figure}[tbp]
\centering
\includegraphics[width=8cm]{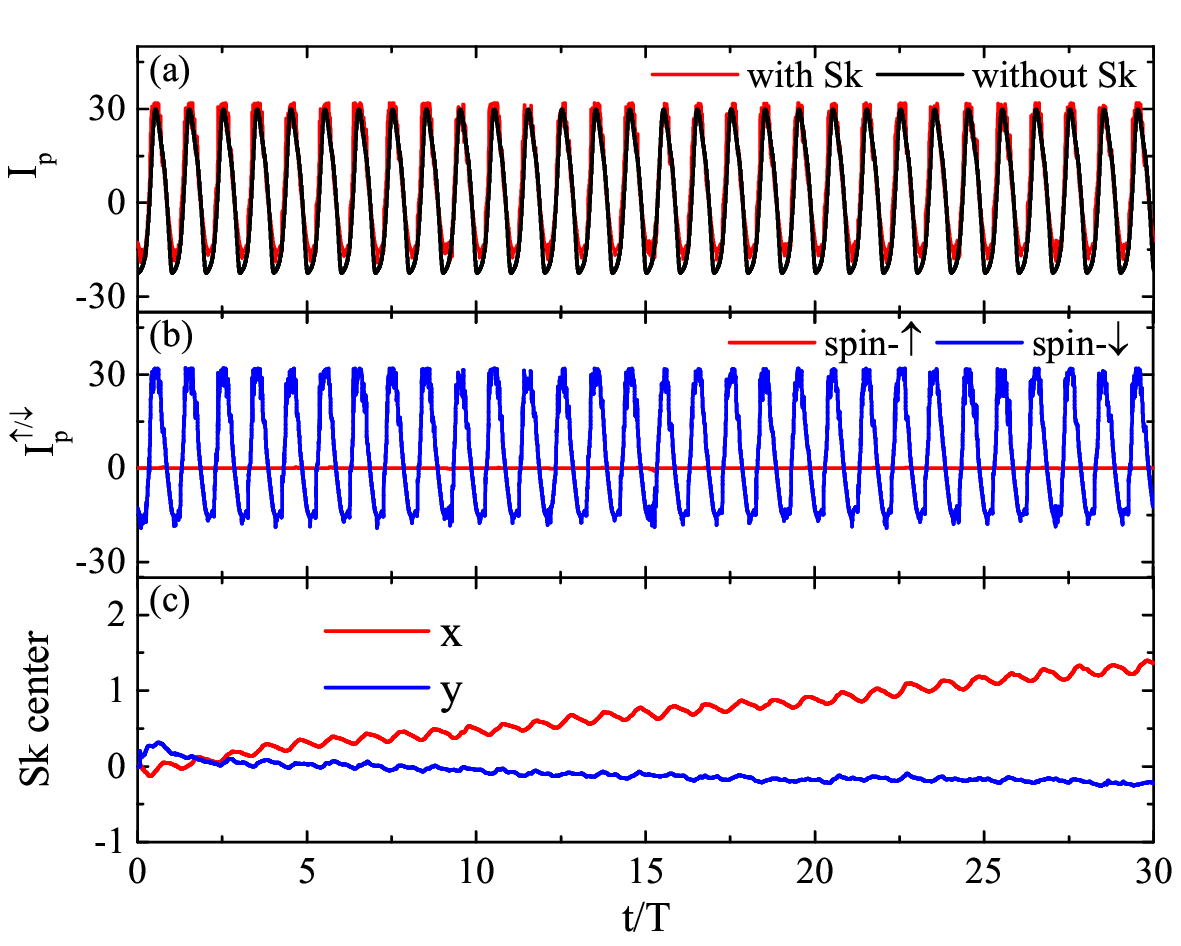}%
\caption{(a) The pumped current $I_{p}$ versus time with or without the Sk. The time is in unit of the pumping period $T$, with $T=2\pi/\omega_p$. (b) The pumped spin-dependent current $I^{\uparrow/\downarrow}_{p}$ with the Sk. (c) Time evolution of the Sk center position $\mathbf{R}=(x, y)$. Parameters: $E=1.9, J=2, V=0.8, \omega_p=1, \phi=\pi/2$.}\label{xy}
\end{figure}

Second, we study the parametric pumping effect on the dynamics of an Sk and the corresponding pumped current. Physically, the pumped current can drive the motion of Sk, while the Sk's motion can affect the pumped current in turn. The pumped current at time $t$ is defined as\cite{Wang-BG2}
\begin{equation}
\label{eq:1}
I_p(t) = {\rm Tr} \left[ \Gamma_R G_f^r \frac{dV_p}{dt} G_f^a \right],
\end{equation}
where $\Gamma_R=\Sigma_R^r-\Sigma_R^a$ is the linewidth function of the right metallic lead. $\Sigma_R^{r,a}$ are the retarded and advanced self-energies. The Sk center $\mathbf{R}=(x, y)$ is defined as $\mathbf{R}=\sum_i(m_0^z-m_i^z)\mathbf{r}_i/\sum_i(m_0^z-m_i^z)$ to characterize its motion, where index $i$ sums over sites with $m_i^z<m_0^z=-0.1$.

As shown in Fig.\ref{xy}(a), in the absence of an Sk, the pumped current is roughly a sine or cosine function in time. When an Sk is introduced at $t=t_0$, the conduction electrons are scattered by the moving Sk. This results in the deviation of the pump current from the smooth curve. Meanwhile, in the presence of the Sk, the pumped current is fully spin-polarized at the given energy, where only the spin-down component is nonzero in Fig.\ref{xy}(b). At the same time, the Sk is driven by the pumped current. Fig.~\ref{xy}(c) displays $x$ and $y$ coordinates of the Sk center as a function of time. The remarkable characteristic is the quasi-periodic movement of the Sk along both $\hat{x}$ and $\hat{y}$ directions. Moreover, the motion of the Sk center has the same period as the pumped current, but is delayed by one quarter cycle in phase. In our system, the pumped current flows in the $\pm x$ direction, and hence the Sk moves faster in this direction. Besides, the Sk acquires a velocity in the $\pm \hat{y}$ direction. This indicates that the Sk Hall effect can also be driven by parametric pumping.

\begin{figure}[tbp]
\centering
\includegraphics[width=8cm]{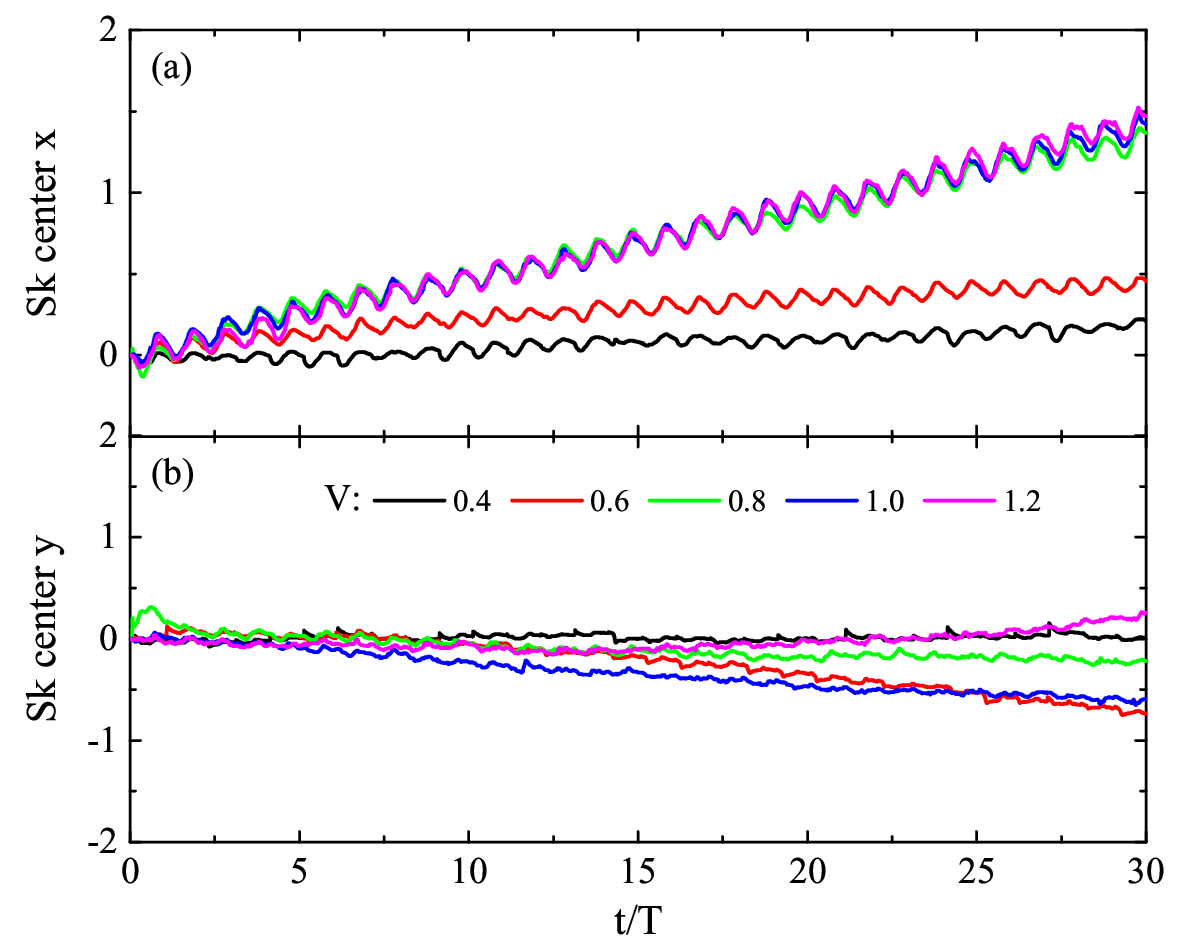}%
\caption{(a) and (b): $x$ and $y$ coordinates of the Sk center $\mathbf{R}$ versus time for different pumping amplitudes $V=0.4, 0.6, 0.8, 1, 1.2$. The pumping frequency is fixed at $\omega_p=1$. Other parameters: $J=2, E=1.9, l_1=1, l_2=5, \phi=\pi/2$.}\label{v0}
\end{figure}

We examine the influence of pumping parameters on the Sk dynamics. The pumping amplitude is first evaluated. Fig.~\ref{v0} shows the time evolution of the Sk center for different pumping amplitudes $V$. We observe that the Sk's speed along $x$ direction increases with the pumping amplitudes. For $V=0.4$, the Sk oscillates around its initial position and does not propagate. As the pumping amplitude increases, the Sk moves faster in $+\hat{x}$ direction and then saturates when the amplitude exceeds $V=0.8$. The motion along the $\hat{y}$ direction is always slower than that in the $\hat{x}$ direction.

\begin{figure}[tbp]
\centering
\includegraphics[width=8cm]{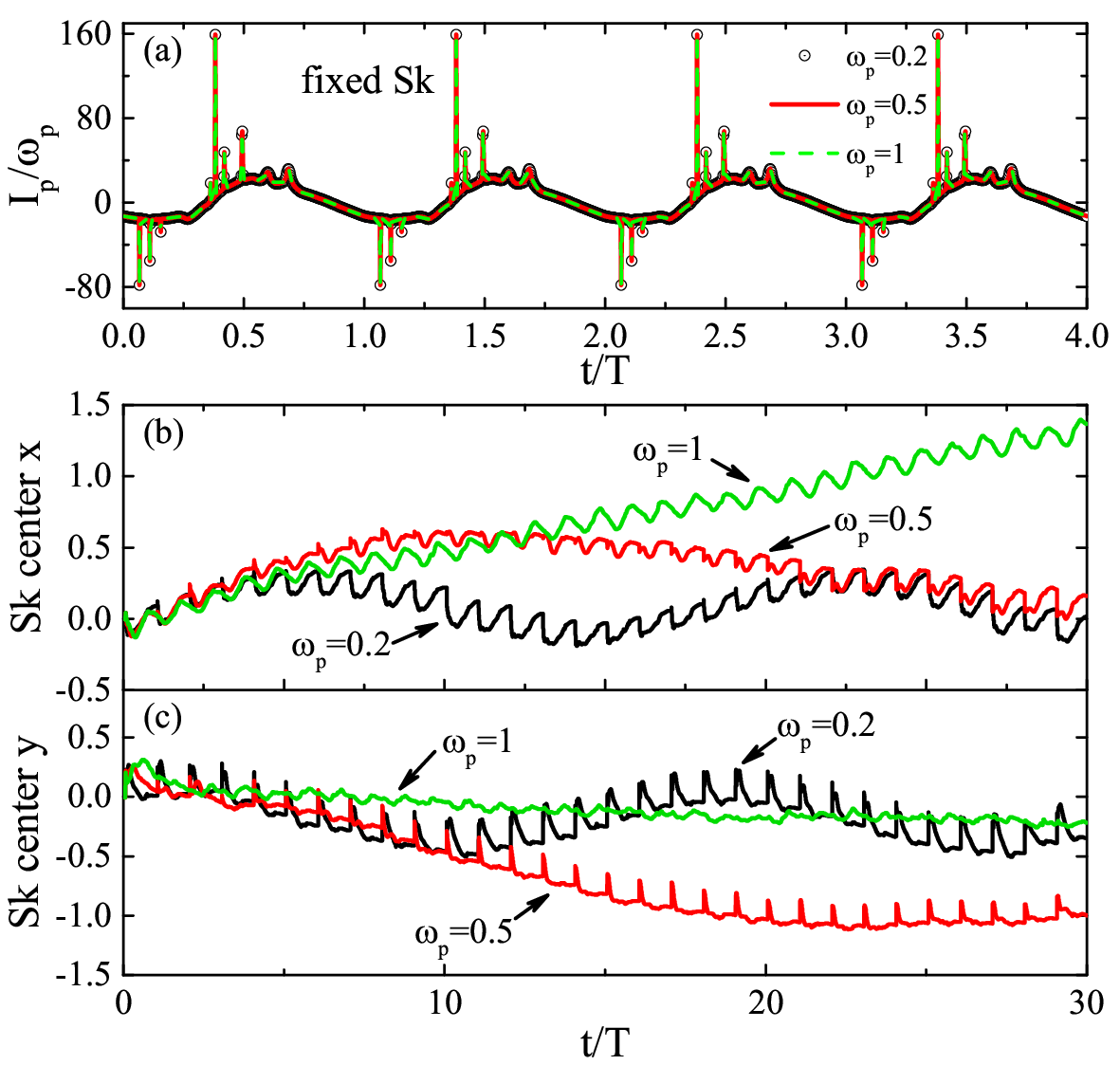}%
\caption{(a) The pumped current for different pumping frequencies. The Sk is fixed at its initial configuration during the time evolution. The pumped current $I_p$ is scaled with $\omega_p$. (b) and (c): $x$ and $y$ coordinates of the Sk center for $\omega_p= 0.2,0.5,1$. The pumping amplitude is fixed at $V=0.8$. Other parameters: $J=2, E=1.9, l_1=1, l_2=5, \phi=\pi/2$. \label{omega}}
\end{figure}

The effect of the pumping frequency is also studied. When there is no Sk, the pumped current in adiabatic pumping regime is independent of the pumping frequency.\cite{brouwer,Wei-YD} In the presence of an Sk, We expect that the pumped current can be simply scaled by the pumping frequency. We examine the pumped current for different pumping frequency when the Sk is fixed at its initial configuration, and numerical results are presented in Fig.~\ref{omega}(a). Four periods are shown here. It is clear that the pumped currents collapse precisely onto each other when scaled by the corresponding pumping frequencies. For a free Sk, Fig.~\ref{omega}(b) and (c) show the $x$ and $y$ coordinates of the Sk center under the pumping. At small frequencies $\omega_p=0.2$ and $0.5$, the Sk is driven along $+\hat{x}$ direction first, and then reflected back periodically. Its motion in $y$ direction is similar. For a larger frequency $\omega_p=1$, there is no such oscillating behavior even for a time scale of 100 periods (not shown here). Notice that when the phase difference is reversed, both the pumped current and the Sk motion change direction.


\section{\label{concludesec}Summary}

In conclusion, we have developed a unified microscopic theory of the LLG equation in terms of nonequilibrium Green's function. Four limiting cases of the microscopic LLG equation are discussed in detail, including the adiabatic, inertia, and nonadiabatic limits for the magnetization with fixed magnitude, as well as the adiabatic limit in spatial domain for the magnetization with slow varying magnitude and direction in space. As a demonstration, the microscopic LLG equation is applied to investigate the motion of a Skyrmion state driven by quantum parametric pumping. Our work not only provides a unified microscopic theory of the LLG equation, but also offers a practical formalism to explore magnetization dynamics with nonequilibrium Green's functions.

\section*{Acknowledgement}
This work was supported by the National Natural Science Foundation of China (Grants No. 12034014, No. 12174262, No. 12074230, and No. 12074190). L. Zhang thanks the Fund for Shanxi "1331 Project".

\section*{Appendix A}
In this appendix, we express the charge and spin current in terms of the nonequilibirum Green's functions.

In the presence of time-varying magnetization, the nonequilibrium Green's function $G^r(t,t')$ depends on two time indices $t$ and $t'$. If the magnetization changes slowly with time, we can treat the time difference $(t-t')$ in energy space.\cite{arrachea2006} After taking the Fourier transformation, the Green's function in energy space only depends on one time variable $t$:
\begin{equation}
G^r(t,E) = \int d\tau e^{iE\tau}G^r(t,t'),\nonumber
\end{equation}
where $\tau = t - t'$. The inverse Fourier transformation gives
\begin{equation}
G^r(t,t') = \int\frac{dE}{2\pi}e^{-iE\tau} G^r(t,E). \nonumber
\end{equation}
With the above definition, it is easy to show that
\begin{equation}
G^{<}(t,t') = \int \frac{dE}{2\pi}e^{-iE\tau} G^r(t,E)\Sigma^{<}(E)G^a(E,t'). \label{lesser}
\end{equation}
In this representation, the particle current matrix is defined\cite{Zhang-L}
\begin{equation}
I^\alpha_{\rm op}(t) = \int\frac{dE}{2\pi}[G^r(t,E)\Sigma_{\alpha}^{<}(E) +G^<(t,E)\Sigma_{\alpha}^a(E) + \rm{H.c.}]. \label{cur-p}
\end{equation}
In term of which, the charge current $I_{c\alpha}(t)$ and spin current $I_{s\alpha}(t)$ are expressed as
\begin{equation}
I_{c\alpha}(t) = -q\mathrm{Tr}[I^\alpha_{\rm op}],~~~{\mathbf{I}}_{s\alpha}(t) = -\frac{1}{2}\mathrm{Tr}[\bm{\sigma} I^\alpha_{\rm op}]. \label{Ic}
\end{equation}

As shown in Appendix C, these time-dependent Green's functions, such as $G^r(t,E)$ (also called Floquet Green's function\cite{arrachea2006}), can be expressed in terms of the instantaneous frozen Green's function $G^r_f$, which satisfies the following recursive relation:
\begin{equation}
G^r(t,E) = G^r_f(t,E) - i G^r_f(t,E)\dot{G}^r(t,E), \label{GF4}
\end{equation}
where $\dot{G}^r$ is the time derivative of $G^r$. The frozen Green's function contains the effective magnetic field ${\bf b}(t)$ and is defined as
\begin{equation}
G_f^{r}(t,E) = [E- H_0 - \Sigma^r - {\bm \sigma}\cdot{\bf b}(t)/2] ^{-1}, \label{GF6}
\end{equation}
where ${\bf b}(t) =\gamma_e\mathbf{B}+J{\mathbf{M}}$. The self energies (for ferromagnetic leads) are given by
\begin{equation}
\Sigma^\gamma_{mn}(t)= \hat{R}^{\dagger} \Sigma^{\gamma}_{0,mn} \hat{R}, \nonumber
\end{equation}
with $\gamma=r,a,<$. $\hat{R}$ is defined in Eq.~(\ref{tunnel02}). $\Sigma^{\gamma}_{0,mn}$ is the self-energy when the magnetization is along $z$-axis:
\begin{equation}
\Sigma^{\gamma}_{0,mn} = \sum_{k\alpha}t^{*}_{k\alpha m} g^{\gamma}_{k\alpha} t_{k\alpha n} =
\begin{pmatrix}
      \Sigma^{\gamma}_{mn,\uparrow} &  0 \\
      0 &  \Sigma^{\gamma}_{mn,\downarrow}
\end{pmatrix}, \nonumber
\end{equation}
where $g^{\gamma}_{k\alpha}$ is the surface Green's function of lead $\alpha$.

\section*{Appendix B}
In this appendix, we express the microscopic LLG equation Eq.~(\ref{dynamic03}) in terms of the spin current and spin operators.

For the spin transfer torque (STT) to occur in magnetic multilayers, one requires a pair of FM layers in a noncollinear configuration. Then a spin-polarized current can be generated from the reference (fixed) layer, and the transverse spin can be transferred to the switchable (free) layer. The current induced STT can also be obtained from Eq.~(\ref{dynamic03}) in such a noncollinear magnetic system. Denoting $H_1 = H_{\rm total}-H'$ and defining the spin operators for lead $\alpha$ and the QD:
\begin{equation}
\hat{\mathbf{s}}_{\alpha} = \frac{1}{2} \sum_{k\sigma\sigma^{'}}\hat{c}^{\dagger}_{k\sigma\alpha}
\bm{\sigma}_{\sigma\sigma^{'}}\hat{c}_{k\sigma^{'}\alpha} \ ,~~~ \hat{\mathbf{s}}_{D}=\frac{1}{2} \sum_{n\sigma\sigma^{'}}\hat{d}^{\dagger}_{n\sigma} \bm{\sigma}_{\sigma\sigma^{'}}\hat{d}_{n\sigma^{'}},\nonumber
\end{equation}
we have
\begin{eqnarray}
\frac{d\hat{\mathbf{s}}_{\alpha}}{dt}  &=& -i[\hat{\mathbf{s}}_{\alpha},H_1], \nonumber \\
\frac{d\hat{\mathbf{s}}_D}{dt} &=& -i[\hat{\mathbf{s}}_D,\hat{H}_1]-i[\hat{\mathbf{s}}_D,\hat{H}'], \nonumber
\end{eqnarray}
where
\begin{equation}
\begin{split}
-i[\hat{\mathbf{s}}_D,\hat{H}']  =
-\gamma_e \hat{\mathbf{s}}_D\times\mathbf{B} + J\hat{\mathbf{M}} \times \hat{\mathbf{s}}_D.
\end{split} \nonumber
\end{equation}
It can be shown that\cite{wangj2006} $[\sum_\alpha \hat{\mathbf{s}}_{\alpha}+\hat{\mathbf{s}}_D,H_1]=0$, from which the spin continuity equation of the system is expressed as\cite{hattori2007}
\begin{equation}
\sum_{\alpha}\frac{d\hat{\mathbf{s}}_{\alpha}}{dt} + \frac{d\hat{\mathbf{s}}_D}{dt} = -\gamma_e \hat{\mathbf{s}}_D\times\mathbf{B}
+J\hat{\mathbf{M}} \times \hat{\mathbf{s}}_D, \label{continuity01}
\end{equation}
which indicates that the total spin is not conserved due to spin precession\cite{wangj2006}. Substituting Eq.~(\ref{continuity01}) into Eq.~(\ref{dynamic01}) and taking quantum average, we have
\begin{equation}
\dot{\mathbf{M}} = -\gamma{\mathbf{M}}\times\mathbf{B} -\sum_{\alpha} {\mathbf I}_{s\alpha} + {\mathbf{A}}, \label{dynamic04}
\end{equation}
where the correction term ${\mathbf{A}}$ is given by
\begin{equation}
{\mathbf{A}} =-\frac{d{\mathbf{s}}_D}{dt}-\gamma_e{\mathbf{s}}_D\times\mathbf{B}, \label{ASD}
\end{equation}
and $\sum_{\alpha}{\mathbf{I}}_{s \alpha} = \sum_{\alpha} d{\mathbf{s}}_{\alpha}/dt$ is the total spin current. Notice that Eq.~(\ref{dynamic03}) and Eq.~(\ref{dynamic04}) are equivalent. Form Eq.~(\ref{ASD}), it is found that when ${\mathbf{s}}_D$ is time-reversal symmetric, $d{\mathbf{s}}_{D}/dt$ or ${\mathbf{A}}$ is time-reversal antisymmetric.
If we decompose $\sum_{\alpha}{\mathbf{I}}_{s\alpha}$ and ${\mathbf{A}}$ into the time-reversal symmetric (labeled with superscript $s$) and antisymmetric (labeled with superscript $a$) parts, we have
\begin{equation}
\dot{\mathbf{M}}=-\gamma{\mathbf{M}}\times\mathbf{B}-\sum_{\alpha}{\mathbf I}^s_{s\alpha} + {\mathbf{A}}^s - J\mathbf{M}\times\mathbf{s}^s_D. \label{dynamic05}
\end{equation}
Here we have used the fact that $\sum_{\alpha} {\mathbf I}_{s\alpha}- {\mathbf{A}} = J\mathbf{M}\times\mathbf{s}_D$. When the external magnetic field is strong enough, the electron spin will approximately align with the direction of the field and we may drop the term $-\gamma_e{\mathbf{s}}_D^a\times\mathbf{B}$.

In the adiabatic limit, the term $-d{\mathbf s}^{a}_D/dt$ in in ${\mathbf{A}}^s$ of Eq.~(\ref{dynamic05}) corresponds to $-d{\mathbf s}^{(0)}_D/dt$. Expanding it to the first order in frequency, we find
\begin{eqnarray}
- \frac{d{\mathbf s}^{(0)}_D}{dt} &=& -\frac{i JM}{4}\int\frac{dE}{2\pi}\mathrm{Tr}[G_f^{<}\bm{\sigma}
G_f^r \bm{\sigma}+ G_f^a \bm{\sigma}G_f^{<}\bm{\sigma}] \frac{d{\mathbf m}}{dt}\nonumber \\
&\equiv& -\eta \frac{d{\mathbf m}}{dt},\label{eq33}
\end{eqnarray}
where $\eta$ is a second rank tensor. This term can be absorbed by introducing an effective gyromagnetic ratio $\gamma'=\gamma (1+\eta)^{-1}$.
Therefore, the driving force of magnetization precession originates from the total spin current $\sum_{\alpha}{\mathbf{I}}^{s}_{s\alpha}$, which corresponds to the current induced STT discussed in Ref.~[\onlinecite{Brataas2012}].

\section*{Appendix C}
In this appendix, we derive the relation between Floquet Green's functions and frozen Green's functions. We start with the two-time retarded Green's function defined as\cite{haugBook}
\begin{equation}
[i\frac{\partial}{\partial t_1} - H(t_1)]G^r(t_1,t_2) + \int dt'\Sigma^r(t_1,t')G^r(t',t_2) = \delta(t_1,t_2).\nonumber
\end{equation}

To work in the Floquet Green's function representation, we make the Fourier transform with respect to the fast time scale $\tau=t_1 - t_2$, and obtain
\begin{equation}
\begin{split}
\Big[i\frac{\partial}{\partial t}
& + E - H(t) \Big] G^r(t,E)  \\
& - \int_{-\infty}^{t} dt' e^{iE(t-t')}
\Sigma^r(t,t')G^r(t',E)
= I.
\end{split} \label{GF}
\end{equation}

The last term on the left hand side can be written as
\begin{equation}
\begin{split}
&\int_{-\infty}^{t} dt' e^{iE(t-t')}
\Sigma^r(t,t')G^r(t',E) \\
& = \int\frac{dE'}{2\pi}\int_{-\infty}^{t} dt' e^{i(E'-E)t'}~ e^{i(E-E')t} \Sigma^r(t,E')G^r(t',E) \\
& \approx \int\frac{dE'}{2\pi}\int_{-\infty}^{t} dt' e^{i(E'-E)t'} ~ e^{i(E-E')t} \Sigma^r(t,E')G^r(t,E)  \\
& = \int\frac{dE'}{2\pi} 2\pi \delta(E'-E)~ e^{i(E-E')t} \Sigma^r(t,E')G^r(t,E) \\
& = \Sigma^r(t,E)G^r(t,E),
\end{split} \nonumber
\end{equation}
where $G^r(t',E) = G^r(t - \tau,E) \approx G^r(t,E)$ is used and the fast time variable $\tau = t - t'$ is neglected. Then we can introduce the frozen Green's function $G^r_f(t,E)$ satisfying
\begin{equation}
[E - H(t) - \Sigma^r(t,E) ] G^r_f(t,E) = I. \label{GF1}
\end{equation}
Substituting Eq.~(\ref{GF1}) into Eq.~(\ref{GF}), we have
\begin{equation}
G^r(t,E) = G^r_f(t,E) - iG^r_f(t,E)\dot{G}^r(t,E). \label{GF2}
\end{equation}
Similarly, the advanced Green's function is
\begin{equation}
G^a(E,t) = G^a_f(E,t) + i\dot{G}^a(E,t){G}^a_f(E,t). \label{GF3}
\end{equation}
Up to the linear order in frequency, we find
\begin{eqnarray}
G^r(t,E) &=& G^r_f(t,E) - iG^r_f(t,E)\dot{G}^r_f(t,E),\nonumber \\
G^a(E,t) &=& G^a_f(E,t) + i\dot{G}^a_f(E,t){G}^a_f(E,t). \label{GF5}
\end{eqnarray}
Notice that the frozen Green's function is an instantaneous function, since it depends only on the present time.

\section*{\label{app:B}Appendix D}

In this appendix, we compare the Gilbert damping coefficient derived in Ref.~[\onlinecite{brataas2008}] using the scattering matrix theory with that obtained in this work via nonequilibrium Green's functions. According to Ref.~[\onlinecite{brataas2008}], in the absence of external bias, the Gilbert tensor element at zero temperature is expressed in terms of the scattering matrix at the Fermi energy:
\begin{equation}
G_{ij}(\mathbf{m}) = \frac{\gamma^2}{4\pi} \mathrm{Re}\{ \mathrm{Tr} [ \frac{\partial S}{\partial m_i} \frac{\partial S}{\partial m_j} ] \}. \label{eq73}
\end{equation}
To compare with Eq.~(\ref{gilbert01}), we calculate the dimensionless quantity $\alpha'_{ij} = G_{ij}/\gamma^2$. The scattering matrix $S$ is connected to the Green's function through the Fisher-Lee relation\cite{fisher1981,wangj2009}
\begin{eqnarray}
S_{\alpha\beta\sigma\sigma'} &=& -\delta_{\alpha\beta\sigma\sigma'} +i\langle W_{\alpha\sigma}|G^r|W_{\beta\sigma'}\rangle, \nonumber \\
(S^\dagger)_{\beta\alpha\sigma'\sigma} &=& -\delta_{\alpha\beta\sigma\sigma'} -i\langle W_{\beta\sigma'}|G^a|W_{\alpha\sigma}\rangle, \nonumber
\end{eqnarray}
where $\Gamma_{\alpha mn \sigma\sigma'} = |W_{\alpha m \sigma}\rangle\langle W_{\alpha n\sigma'}|$. $|W_{\alpha m}\rangle$ is proportional to the eigenvector of $\Gamma_{\alpha}$\cite{wangj2009}. From Eq.~(\ref{GF6}), we see that
\begin{equation}
\frac{\partial G^{r,a}_f}{\partial m_i} =\frac{1}{2}JM G^{r,a}_f \sigma_i G^{r,a}_f. \nonumber
\end{equation}
Then we can express Eq.~(\ref{eq73}) with frozen Green's functions
\begin{eqnarray}
\mathrm{Tr}[\frac{\partial S}{\partial m_i}\frac{\partial S}{\partial m_j}]&=&\frac{1}{4}(JM)^2\mathrm{Tr}[G^a_f \Gamma G^r_f \sigma_i G^r_f\Gamma G^a_f \sigma_j]\nonumber \\
&=& (JM)^2 \mathrm{Tr} [{\rm Im}(G^r_f) \sigma_i {\rm Im}(G^r_f) \sigma_j ]. \nonumber
\end{eqnarray}
Hence the dimensionless damping tensor element is
\begin{equation}
\alpha'_{ij} = \frac{1}{16\pi}(JM)^2 \mathrm{Re}\{ \mathrm{Tr}[ G^a_f \Gamma G^r_f \sigma_i G^r_f \Gamma G^a_f \sigma_j ] \}. \label{gilbert03}
\end{equation}
Now we assume that there is no external bias and the temperature is zero, which are the same conditions used in deriving Eq.~(\ref{eq73}) in Ref.~[\onlinecite{brataas2008}]. Note that Eq.~(\ref{eq73}) was derived from the pumped energy current defined as ${\dot {\bf m}} \alpha' {\dot {\bf m}}$ so that the resultant tensor $\alpha'_{ij}$ is always symmetric. Hence we symmetrize the dimensionless damping tensor in Eq.~(\ref{gilbert01}):
\begin{equation}
\tilde{\alpha}_{ij} = \frac{1}{2} (\alpha_{ij} + \alpha_{ji}). \label{gilbert06}
\end{equation}
In the absence of external bias, $G_f^{<} = (G_f^a - G_f^r) f(E)$, and Eq.~(\ref{gilbert06}) becomes
\begin{eqnarray}
\tilde{\alpha}_{ij} &&= -\int dE \mathrm{Tr} [\sigma_i A \sigma_j G_f^r G_f^r - \sigma_j A \sigma_i G_f^a G_f^a \nonumber \\
&&+ \sigma_j A \sigma_i G_f^r G_f^r - \sigma_i A \sigma_j G_f^a G_f^a] \nonumber \\
&&=-\int dE \mathrm{Tr} [\sigma_i A \sigma_j \partial_E(G_f^r -G^a_f) + \sigma_j A \sigma_i \partial_E(G_f^r -G^a_f)]\nonumber \\
&&=-\int dE f \mathrm{Tr} \partial_E [\sigma_i (G_f^r -G^a_f)\sigma_j (G_f^r -G^a_f)] \nonumber \\
&&= \mathrm{Tr} [\sigma_i (G_f^r -G^a_f)\sigma_j (G_f^r -G^a_f)], \label{aij}
\end{eqnarray}
with $A = (G_f^r -G^a_f)f$. Apart from a constant factor $(JM)^2/16\pi$, this expression is exactly the same as $\alpha'_{ij}$ shown in Eq.~(\ref{gilbert03}). Therefore, we confirm the equivalence of the Gilbert damping tensor between our formalism (Eq.~(\ref{gilbert01})) and that obtained in Ref.~[\onlinecite{brataas2008}] in the limiting case.

\section*{\label{app:C}Appendix E}

In this appendix, we derive the nonadiabatic term in Eq.~(\ref{LLG03}). Expanding ${\bf m}({\bf r},t)$ up to the first order in $\partial_i {\bf m} \equiv \partial {\bf m}/\partial x_i$, we have
\begin{equation}
{\bf m}({\bf r},t)={\bf m}+ \delta {\bf m}={\bf m}+ x_i \partial_i {\bf m}, \nonumber
\end{equation}
where the Einstein summation convention is implied. The nonequilibrium Green's functions have similar expansions
\begin{eqnarray}
G^r &=& G^r_0 - (JM/4) G^r_0 {\bm \sigma} x_i \partial_i {\bf m} G^r_0, \nonumber \\
G^< &=& G^<_0 - [(JM/4) G^r_0 {\bm \sigma} x_i \partial_i {\bf m} G^<_0 - \rm{H.c.}]. \nonumber
\end{eqnarray}
It is straightforward to find the correction on ${\bm s}^{(0)}_D$ due to $\partial_i {\bf m}$,
\begin{equation}
\delta {\bm s}^{(0)}_D = \frac{JM i}{8}\int \frac{dE}{2\pi} {\rm Tr_s}[ {\bm \sigma}G^r {\bm \sigma} x_j G^<]_{x x} \partial_j {\bf m} + \rm{H.c.}, \label{eq47}
\end{equation}
where we have focused on the linear response regime\cite{Zhang-SF,Li-Z} and kept only the linear term in $\partial_j {\bf m}$. Here ${\rm Tr_s}[...]_{xx}$ denotes tracing over spin space and then taking diagonal matrix element in real space and we have dropped the subscript $0$ in the Green's function. If we further neglect SOI, the Green's function is diagonal in spin space in the linear regime, and ${\rm Tr_s}[ {\bm \sigma}G^r {\bm \sigma} x_j G^<]_{x x}={\bm 1} {\rm Tr_s}[G^r x_j G^<]_{x x}$ is also diagonal in spin space. Using the relation
\begin{equation}
x_j = m c_0 [\nabla_j, H-E], \nonumber
\end{equation}
which is valid in the adiabatic approximation in spatial domain ($c_0$ is a constant having dimension $T^{-2}$ with $T$ the dimension of time),  Eq.~(\ref{eq47}) becomes
\begin{equation}
\delta {\bm s}^{(0)}_D = \frac{JM m c_0 }{8}[w_1(x)+w_2(x)] \partial_j {\bf m}, \nonumber
\end{equation}
where
\begin{eqnarray}
w_{1j}(x) &=& i\int \frac{dE}{2\pi}[(G^< \nabla_j ) G^a (H-E) - (H-E) G^r (\nabla_j G^<)], \nonumber \\
w_{2j}(x) &=& i\int \frac{dE}{2\pi}[(G^r \nabla_j ) (H-E) G^< - G^<(H-E) (\nabla_j G^a)]. \nonumber
\end{eqnarray}
Using ${\bm 1}+\Sigma^r G^r = (E-H) G^r$, we find
\begin{eqnarray}
w_{1j}(x) &=& i\int \frac{dE}{2\pi}[(\nabla_j G^<)-(G^< \nabla_j )] \nonumber \\
&+& \int \frac{dE}{2\pi}[\Sigma^r G^r (\nabla_j G^<)-(G^< \nabla_j )G^a \Sigma^a], \nonumber
\end{eqnarray}
where the first term of $w_{1j}(x)$ is proportional to the current density ${\bm j}_e$\cite{Zhang-L2}. Focusing on this particular term, we have
\begin{equation}
\delta {\bm s}^{(0)}_D = \frac{JM m^2 c_0}{4}({\bm j}_e \cdot \nabla) {\bf m}, \nonumber
\end{equation}
which gives rise to the nonadiabatic torque due to the spatial variation of the magnetization.


\begin{thebibliography}{25}
\bibitem{timescale}
In deriving Eq.~(\ref{dynamic02}), we have assumed that $\mathbf{M}$ varies slowly with time and its time scale is much larger than that of electrons. Hence the Born-Oppenheimer approximation can be used to separate the degrees of freedom of the magnetization and electrons. It is reasonable since typical time scale for the magnetization dynamics is in the order of nanoseconds while the electronic time scale is in femtoseconds. For an electric pulse, the time scale can be as short as $ps$ which is still much larger than $fs$. In the limit of slow precession, the magnetization dynamics can be approximated as $\delta\dot{\hat{\mathbf{M}}}\approx 0$. The magnetization spin is treated as a classical vector, which rotates around the magnetic field. We can further neglect the magnetization fluctuation term ($\delta\hat{\mathbf{M}} \approx 0$). Then the fluctuation comes solely from the electron spin, $\delta\mathcal{{\hat B}} = -J ~ \delta\hat{\mathbf{s}}_D$.
\end{thebibliography}
\end{document}